\documentstyle[12pt,axodraw]{article}

\catcode`@=12
\begin{document}
\thispagestyle{empty}
\begin{flushright} 
UCRHEP-T314\\ 
November 2001\
\end{flushright}
\vspace{0.5in}
\begin{center}
{\Large	\bf Unified Supersymmetric Model of\\ Naturally Small Dirac
 Neutrino Masses and\\ the Axionic Solution of the Strong CP Problem\\}
\vspace{2.0in}
{\bf Ernest Ma\\}
\vspace{0.2in}
{\sl Physics Department, University of California, Riverside, 
California 92521\\}
\vspace{2.0in}
\end{center}
\begin{abstract}\
Using the particle content of the fundamental \underline {27} supermultiplet 
of $E_6$, naturally small Dirac neutrino masses are obtained in the context 
of $SU(3)_C \times SU(2)_L \times U(1)_Y \times U(1)_\chi$, where $U(1)_\chi$ 
comes from the decomposition $E_6 \to SO(10) \times U(1)_\psi$, then $SO(10) 
\to SU(5) \times U(1)_\chi$.  New observable consequences are predicted at 
the TeV scale.  An axionic solution of the strong CP problem may be 
included at no extra cost.
\end{abstract}
\newpage
\baselineskip 24pt

With the present experimental evidence \cite{atm,solar,lsnd} on neutrino 
oscillations, the notion that neutrinos should be massive is no longer 
in dispute.  The next question is whether neutrino masses are Majorana 
or Dirac.  Experimentally, the nonobservation of neutrinoless double beta 
decay at the 0.2 eV level \cite{bb} is unable to settle this issue, but 
there are very strong and convincing theoretical reasons to believe that 
neutrino masses should be Majorana.  On the other hand, if the theoretical 
context is changed, naturally small Dirac neutrino masses are possible, 
as shown below.

To obtain a Dirac mass, the left-handed neutrino $\nu_L$ must be paired with 
a right-handed singlet $N_R$.  Two problems arise immediately.  ($i$) There 
is no symmetry to prevent $N_R$ from acquiring a large Majorana mass. 
($ii$) Even if such a symmetry (such as additive lepton number) is imposed, 
an extremely small Yukawa coupling (less than $10^{-11}$) is still needed to 
satisfy the experimental bound $m_\nu < $ a few eV.  The usual resolution of 
these problems is to take advantage of ($i$) to make $m_N$ very large, so that 
the famous canonical seesaw mechanism \cite{seesaw} makes $m_\nu = m_D^2 / 
m_N$. Now ($ii$) is also not a problem because the Yukawa coupling for the 
Dirac mass $m_D$ is no longer required to be very small.

In this paper a new scenario is proposed where ($i$) $N_R$ is naturally 
prevented from having a Majorana mass and ($ii$) $m_D$ is small without 
having a small Yukawa coupling \cite{ma00,ma01}.  This is possible because 
the theoretical framework used will be that of superstring-inspired $E_6$ 
\cite{e6}.  As a bonus, the axionic solution \cite{pq} of the strong CP 
problem may also be included.

The starting point is the gauge group $E_6$ and its decomposition 
$E_6 \to SO(10) \times U(1)_\psi$, then $SO(10) \to SU(5) \times U(1)_\chi$. 
It is often assumed that at TeV energies, a linear combination of $U(1)_\psi$ 
and $U(1)_\chi$ remains \cite{u1} in addition to the standard $SU(3)_C 
\times SU(2)_L \times U(1)_Y$.  It is usually also assumed that three 
complete matter supermultiplets of the fundamental \underline {27} 
representation of $E_6$ are present at these energies, which include the 
known three families of quarks and leptons as well as other new particles. 
Under the subgroup $SU(5) \times U(1)_\psi \times U(1)_\chi$, the particle 
content of each supermultiplet is given by
\begin{eqnarray}
\underline {27} &=& (10;1,-1)[(u,d),u^c,e^c] + (5^*;1,3)[d^c,(\nu_e,e)] + 
(1;1,-5)[N^c] \nonumber \\ &+& (5;-2,2)[h,(E^c,N_E^c)] + (5^*;-2,-2)[h^c,
(\nu_E,E)] + (1;4,0)[S],
\end{eqnarray}
where the U(1) charges refer to $2\sqrt 6 Q_\psi$ and $2 \sqrt {10} Q_\chi$. 
Note that the known quarks and leptons are contained in $(10;1,-1)$ and 
$(5^*;1,3)$, and the two Higgs scalar doublets are represented by $(\nu_E,E)$ 
and $(E^c,N_E^c)$.  Since $N^c$ and $S$ are singlets under $SU(5)$, one 
linear combination will be trivial under the assumed low-energy gauge group, 
i.e. $SU(3)_C \times SU(2)_L \times U(1)_Y \times U(1)_\alpha$ with 
$Q_\alpha = Q_\psi \cos \alpha + Q_\chi \sin \alpha$.  For the choice 
$\tan \alpha = \sqrt {1/15}$, the $U(1)_N$ model \cite{ma96,hmrs} is obtained, 
for which $N^c$ is trivial, thus allowing it to acquire a large Majorana 
mass.  Combining this with the existing term $(\nu_e N_E^c - e E^c) N^c$, 
the usual seesaw Majorana neutrino mass may then be obtained.

Consider now the case $\sin \alpha = 1$, i.e. the $U(1)_\chi$ model.  This 
allows $S$ to have a large Majorana mass, but not $N^c$.  Hence the only 
apparent way that $\nu_e$ may become massive is to pair up with $N^c$ to 
form a Dirac neutrino with mass proportional to the vacuum expectation value 
(VEV) of the scalar component of $N_E^c$.  If the latter is of the order of 
the electroweak symmetry breaking scale, i.e. $10^2$ GeV, then an extremely 
small Yukawa coupling is required.  This is in fact the prevailing working 
ansatz of all $U(1)_\alpha$ models except $U(1)_N$.  However, there is a 
very simple and natural solution.  If $\tilde N_E^c$ has $m^2 > 0$ with 
$m$ large, then its VEV can be very small \cite{ma00,ma01}.  This is 
precisely the case in the $U(1)_\chi$ model, where $\nu_E N_E^c - E E^c$ is 
an allowed term.

There are 11 generic terms \cite{ma88} in the superpotential of such $E_6$ 
models.  They are
\begin{eqnarray}
(1) && \hat Q \hat u^c \hat {\bar E} = (\hat u \hat N_E^c - \hat d \hat E^c) 
\hat u^c, \\ 
(2) && \hat Q \hat d^c \hat E = (\hat u \hat E - \hat d \hat \nu_E) \hat d^c, 
\\ 
(3) && \hat L \hat e^c \hat E = (\hat \nu_e \hat E - \hat e \hat \nu_E) \hat 
e^c, \\ 
(4) && \hat S \hat E \hat {\bar E} = (\hat \nu_E \hat N_E^c - \hat E \hat E^c) 
\hat S, \\ 
(5) && \hat S \hat h \hat h^c, \\
(6) && \hat L \hat N^c \hat {\bar E} = (\hat \nu_e \hat N_E^c - \hat e \hat 
E^c) \hat N^c, \\ 
(7) && \hat Q \hat L \hat h^c = (\hat u \hat e - \hat d \hat \nu_e) \hat h^c, 
\\ 
(8) && \hat u^c \hat e^c \hat h, \\ 
(9) && \hat d^c \hat N^c \hat h, \\ 
(10) && \hat Q \hat Q \hat h = (\hat u \hat d - \hat d \hat u) \hat h, \\ 
(11) && \hat u^c \hat d^c \hat h^c.
\end{eqnarray}
To prevent rapid proton decay, some terms must be absent.  This is usually 
accomplished by the imposition of an exactly conserved discrete symmetry, 
such as the well-known $R$ parity.  Here the choice is
\begin{equation}
Z_3 \times U(1)_{PQ}.
\end{equation}
Under $Z_3$ with $\omega^3 = 1$, $\hat E_{1,2}$, $\hat {\bar E}_{1,2}$ 
transform as $\omega$; $\hat u^c$, $\hat d^c$, $\hat e^c$ as $\omega^2$; 
and all other superfields as 1.  Under $U(1)_{PQ}$, the only superfields with 
nonzero charges are $\hat h, \hat h^c, \hat S_1, \hat S_2, \hat S_3$ with 
charges $1/2, 1/2, -1, 2, -2$ respectively.  This means that the terms (7) 
to (11) are all forbidden, the term (6) involves only $\hat {\bar E}_3$, the 
term (5) involves only $\hat S_1$, and the term (4) is forbidden, but since 
$\hat S$ is trivial under $U(1)_\chi$, the soft term
\begin{equation}
\hat E \hat {\bar E} = \hat \nu_e \hat N_E^c - \hat E \hat E^c
\end{equation}
by itself is allowed.  Note that only one term, i.e. $\hat E_3 \hat 
{\bar E}_3$, is invariant under $Z_3$.  All other $\hat E \hat {\bar E}$ 
terms will break $Z_3$ but only \underline {softly}.

The superpotential of this model is then given by
\begin{eqnarray}
\hat W &=& \mu_{ij} \hat E_i \hat {\bar E}_j + f^{(u)}_{(1,2)ij} \hat Q_i 
\hat u^c_j \hat {\bar E}_{1,2} + f^{(d)}_{(1,2)ij} \hat Q_i \hat d^c_j \hat 
E_{1,2} + f^{(e)}_{(1,2)ij} \hat L_i \hat e^c_j \hat E_{1,2} \nonumber \\ 
&+& f^{(N)}_{ij} \hat L_i \hat N^c_j \hat {\bar E}_3 + f^{(h)}_{ij} \hat S_1 
\hat h_i \hat h^c_j + m_2 \hat S_2 \hat S_3 + f_2 \hat S_2 \hat S_1 \hat S_1.
\end{eqnarray}
The anomalous global $U(1)_{PQ}$ is spontaneously broken at the intermediate 
scale $m_2$ so that an ``invisible'' axion will emerge to solve the strong CP 
problem.  The $U(1)_{PQ}$ charges of $\hat S_{1,2,3}$ are chosen so that 
$S_1$ may acquire a large VEV $(\sim m_2 \sim 10^9$ to $10^{12}$ GeV) 
without breaking the 
supersymmetry of the entire theory at that scale.  Details are contained in 
Ref.\cite{ma01_axion}.  Because the usual quarks and leptons here do not 
transform under $U(1)_{PQ}$, the axion of this model is of the KSVZ type 
\cite{ksvz}, whereas that of Ref.\cite{ma01_axion} is of the DFSZ type 
\cite{dfsz}. 

Note that $U(1)_{PQ}$ here serves the \underline {dual} purpose of solving 
the problem of rapid proton decay as well.  Note also that the choice of 
$U(1)_\chi$ as the extra gauge symmetry is the only one which allows that to 
work.  It also serves the purpose of allowing the term $\hat 
E \hat {\bar E}$ and the choice of $Z_3$ allows only $\hat {\bar E_3}$ to 
couple to $\hat N^c$, with a large mass for $\hat E_3 \hat {\bar E}_3$.  The 
Dirac mass linking $\nu_e$ to $N^c$ is proportional to the VEV of the scalar 
component of $\hat {\bar E}_3$, which may then be very small \cite{ma00,ma01}, 
as shown below. 

Consider the following Higgs potential of 4 scalar doublets $H_{1,2,3,4}$ 
representing the scalar components of $\hat E_1, \hat {\bar E}_1, \hat E_3, 
\hat {\bar E}_3$ respectively \cite{ma01_comb} (assuming that $\hat E_2$ and 
$\hat {\bar E}_2$ have no VEV):
\begin{eqnarray}
V &=& \sum_i m_i^2 H_i^\dagger H_i + [m_{13}^2 H_1^\dagger H_3 + m_{24}^2 
H_2^\dagger H_4 \nonumber \\ 
&& + ~m_{12}^2 H_1 H_2 + m_{14}^2 H_1 H_4 + 
m_{32}^2 H_3 H_2 + m_{34}^2 H_3 H_4 + h.c.] \nonumber \\ 
&+& {1 \over 2} \left( {g_1^2 \over 4} + {g_\chi^2 \over 10} \right) 
[-H_1^\dagger H_1 + H_2^\dagger H_2 - H_3^\dagger H_3 + H_4^\dagger H_4]^2 
\nonumber \\ 
&+& {1 \over 2} g_2^2 \sum_\alpha |\sum_i H_i^\dagger \tau_\alpha 
H_i|^2,
\end{eqnarray}
where $\tau_\alpha (\alpha = 1,2,3)$ are the usual SU(2) representation 
matrices.  Let the VEV's of $H_i$ be $v_i$, then the minimum of $V$ is
\begin{eqnarray}
V_{min} &=& \sum_i m_i^2 v_i^2 + 2 m_{12}^2 v_1 v_2 + 2 m_{13}^2 v_1 v_3 + 
2 m_{14}^2 v_1 v_4 + 2 m_{24}^2 v_2 v_4 + 
2 m_{32}^2 v_2 v_3 + 2 m_{34}^2 v_3 v_4 \nonumber \\ 
&+& {1 \over 8} \left( g_1^2 + g_2^2 + {2 g_\chi^2 \over 5} \right) 
(v_1^2 - v_2^2 + v_3^2 - v_4^2)^2,
\end{eqnarray}
where all parameters have been assumed real for simplicity.  The 4 equations 
of constraint are
\begin{eqnarray}
0 &=& m_1^2 v_1 + m_{12}^2 v_2 + m_{13}^2 v_3 + m_{14}^2 v_4 + {1 \over 4} 
\left( g_1^2 + 
g_2^2 + {2 g_\chi^2 \over 5} \right) v_1 (v_1^2 - v_2^2 + v_3^2 - v_4^2), \\ 
0 &=& m_2^2 v_2 + m_{12}^2 v_1 + m_{24}^2 v_4 + m_{32}^2 v_3 - {1 \over 4} 
\left( g_1^2 + 
g_2^2 + {2 g_\chi^2 \over 5} \right) v_2 (v_1^2 - v_2^2 + v_3^2 - v_4^2), \\ 
0 &=& m_3^2 v_3 + m_{13}^2 v_1 + m_{32}^2 v_2 + m_{34}^2 v_4 + {1 \over 4} 
\left( g_1^2 + 
g_2^2 + {2 g_\chi^2 \over 5} \right) v_3 (v_1^2 - v_2^2 + v_3^2 - v_4^2), \\ 
0 &=& m_4^2 v_4 + m_{24}^2 v_2 + m_{14}^2 v_1 + m_{34}^2 v_3 - {1 \over 4} 
\left( g_1^2 + 
g_2^2 + {2 g_\chi^2 \over 5} \right) v_4 (v_1^2 - v_2^2 + v_3^2 - v_4^2).
\end{eqnarray}
Since $m_3^2 \sim m_4^2 \sim \mu_{33}^2$, $m_{13}^2 \sim \mu_{13} 
\mu_{33}$, and $m_{24}^2 \sim \mu_{31} \mu_{33}$ are the only parameters 
which have contributions involving the large mass $\mu_{33}$, it is clear 
that Eqs.~(20) and (21) have the solution
\begin{equation}
v_3 \simeq -{m_{13}^2 v_1 \over m_3^2}, ~~~ v_4 \simeq -{m_{24}^2 v_2 \over 
m_4^2}.
\end{equation}
They may then be of order 0.1 eV if $m_{3,4} \sim \mu_{33} \sim 10^{15}$ GeV 
(i.e. close to a possible grand-unification mass scale), 
and $\mu_{13}, \mu_{31} \sim M_{SUSY} \sim$ 1 TeV.  Setting $v_3 = v_4 
= 0$ in Eqs.~(18) and (19), the usual conditions of the minimal supersymmetric 
standard model are obtained except for the additional terms due to $g_\chi$. 

Now $U(1)_\chi$ also undergoes spontaneous symmetry breaking through the 
VEV of one linear combination of the 3 $(\tilde N^c)$'s.  As a result, there 
appear a new massive neutral gauge boson $Z'$, the corresponding scalar boson 
$\sqrt 2 Re \tilde N^c$, and the Dirac fermion which comes from the pairing 
of $\tilde z'$ and $N^c$, all having the mass $(\sqrt 5/2) g_\chi \langle 
\tilde N^c \rangle$ \cite{km}.  Hence only 2 $(N^c)$'s remain and they combine 
with 2 of the 3 $\nu$'s to form 2 light Dirac neutrinos.  The remaining 
$\nu$ gets a negligible Majorana mass from the allowed supersymmetry-breaking 
soft Majorana mass of $\tilde z'$.  A satisfactory framework is thus 
established for describing the oscillations of 2 light Dirac neutrinos 
and 1 essentially massless Majorana neutrino.

At the TeV energy scale, this model is verifiable experimentally by its 
many unique predictions.  First, there must be a $Z'$ gauge boson with 
couplings to quarks and leptons according to Eq.~(1).  In particular, it 
will have invisible decays to neutrinos given by
\begin{equation}
{\Gamma (Z' \to \bar \nu \nu + \bar N^c N^c) \over \Gamma (Z' \to l^+ l^-)} 
= {77 \over 30}.
\end{equation}
There are likely to be 4 Higgs doublets, instead of 2, and definitely not 6.  
There should not be exotic quarks (i.e. $h$ and $h^c$) because they are 
predicted to be very heavy with masses at the axion scale.  The axion itself 
is of course very light and very difficult to detect \cite{rvb}.  Its 
partners, the saxion and the axino, are likely to be at or below the TeV 
scale and may also be components of the dark matter of the Universe. 
Lepton number is violated through $\langle \tilde N^c \rangle$, but 
since $\hat N^c$ only appears in Eq.~(15) with the very heavy 
$\hat {\bar E}_3$, this violation is highly suppressed.  Thus my proposed 
model evades the general conclusion of Ref.\cite{hmrs} regarding $E_6$ 
subgroups that only $U(1)_N$ \cite{ma96} and the skew left-right model 
\cite{skew} do not have lepton-number violating interactions at the TeV scale 
which would erase any preexisting lepton or baryon asymmetry of the Universe.

In conclusion, a new unified supersymmetric model has been proposed which has 
the following desirable properties.  

(1) Its particle content comes from 3 
complete fundamental \underline {27} representations of $E_6$, which may be 
the remnant of an underlying superstring theory.  

(2) Its low-energy gauge 
group is $SU(3)_C \times SU(2)_L \times U(1)_Y \times U(1)_\chi$, where 
$U(1)_\chi$ comes from $E_6 \to SO(10) \to SU(5) \times U(1)_\chi$.  

(3) It 
has the additional symmetry $Z_3 \times U(1)_{PQ}$ which serves many purposes, 
including that of preventing rapid proton decay.  $Z_3$ is softly broken; 
$U(1)_{PQ}$ is spontaneously broken.  

(4) Naturally small Dirac neutrino 
masses \cite{other,extra} come from the $\hat L \hat N^c \hat {\bar E}_3$ 
term of Eq.~(15) because $\tilde {\bar E}_3$ has a very small VEV, using the 
mechanism \cite{ma00,ma01} of a large positive $m^2$ close to a possible 
grand-unification mass scale for $\tilde {\bar E}_3$, 
as shown by Eq.~(22).  

(5) The 3 singlet superfields $\hat S_{1,2,3}$, which 
do not transform under $U(1)_\chi$, are chosen \cite{ma01_axion} to obtain 
an axionic solution of the strong CP problem, such that $f_a >> M_{SUSY}$. 

(6) This model 
predicts a definite supersymmetric particle structure associated with the 
extra $U(1)_\chi$ gauge symmetry at the TeV scale, which should be accessible 
in near-future high-energy accelerators.  

(7) It is the only model to date 
which incorporates naturally small \underline {Dirac} neutrino masses with 
the axionic solution of the strong CP problem in a comprehensive theoretical 
framework of all particle interactions.

This work was supported in part by the U.~S.~Department of Energy
under Grant No.~DE-FG03-94ER40837.

\bibliographystyle{unsrt}

\end{document}